\documentclass[a4paper,11pt]{article}
\pdfoutput=1 
\usepackage{jheppub}
\usepackage{amsmath}
\usepackage{amssymb}
\usepackage{color}
\usepackage{xcolor}
\usepackage{graphicx}
\usepackage{hyperref}
\usepackage{textcomp}
\usepackage{xspace,slashed}
\usepackage[utf8]{inputenc}
\usepackage{subcaption}
\usepackage{multirow}
\usepackage{nccmath} 
\usepackage{slashed}
\usepackage[bb=boondox]{mathalfa} 
\usepackage{orcidlink}
\usepackage{longtable}

\makeatletter
\renewcommand\@fpheader{}
\renewcommand\@journal{}
\makeatother

\newcommand{\ep}{\epsilon}
\newcommand{\intl}{I}
\newcommand{\intd}{\mathcal{I}}

\title{On the finite basis of two-loop `t Hooft-Veltman Feynman integrals}

\author[a]{Piotr Bargie\l{}a~\orcidlink{0000-0002-3646-5892}}
\emailAdd{piotr.bargiela@physik.uzh.ch}
\author[a]{Tong-Zhi Yang~\orcidlink{0000-0001-5003-5517}}
\emailAdd{tongzhi.yang@physik.uzh.ch}
\affiliation[a]{Physik-Institut, Universit\"at Z\"urich, Winterthurerstrasse 190, 8057 Z\"urich, Switzerland}

\preprint{ZU-TH 15/25}

\abstract{
In this work, we investigate the finite basis topologies of two-loop dimensionally regularized Feynman integrals in the `t Hooft-Veltman scheme in the Standard Model.
We present a functionally distinct finite basis of Master Integrals which spans the whole transcendental space of all two-loop Feynman integrals with external momenta in four dimensions.
We also indicate that all the two-loop Master Integrals, in an appropriate basis, with more than 8 denominators do not contribute to the finite part of any two-loop scattering amplitude.
In addition, we elaborate on the application of the `t Hooft-Veltman decomposition to improve the performance of numerical evaluation of Feynman integrals using AMFlow and DCT packages.
Moreover, we analyze the spectrum of special functions and the corresponding geometries appearing in any two-loop scattering amplitude.
Our work will allow for a reduction in the computational complexity required for providing high-precision predictions for future high-multiplicity collider observables, both analytically and numerically.
}

\begin{document}

\maketitle 

\allowdisplaybreaks

\section{Introduction}
\label{sec:intro}

Feynman integrals are the source of transcendental functions in scattering amplitudes in perturbative Quantum Field Theory of the Standard Model in Particle Physics.
Calculating them is necessary for providing high-precision phenomenological predictions for collider observables.
In practice, this poses a huge challenge, both analytically, and computationally. 
There has been a major recent progress in understanding analytic and numerical properties of Feynman integrals, see an example review in Ref.~\cite{Bourjaily:2022bwx}.
In result, the relevant integrals have been solved for a multitude of scattering processes, see an example review in Ref.~\cite{Heinrich:2020ybq}.

Mathematically, scattering amplitudes depend on three structures, i.e. color, helicity, and kinematics.
After performing the color, Lorentz tensor, and Dirac spinor algebras, the amplitude can be linearly decomposed into a sum of scalar Feynman integrals, which source the transcendental dependence on the external kinematic invariants.
In this work, we will focus on such scalar integrals, with standard quadratic propagators.
In addition, for physical processes, Feynman integrals usually diverge when integrating explicitly in $d=4$ dimensions.
In order to regulate them, we choose here the most common scheme i.e. the dimensional regularization in $d=d_0-2\ep$ around $\ep \to 0$.
We note that, while we proceed here with an analysis for integrals with standard quadratic propagators in $d_0=4$, similar arguments could be applied for integrals arising in other Effective Field Theories and dimensions $d_0$.

It is well known that only 5 propagators can be linearly independent at the one-loop perturbative order in $d_0=4$~\cite{Melrose:1965kb}.
It is because there can be at most 4 independent external momenta in $d_0=4$.
This enables the reduction of any high-multiplicity one-loop integral in terms of a pentagon integral and its subsectors~\cite{Passarino:1978jh,vanNeerven:1983vr,Bern:1992em,Binoth:1999sp,Fleischer:1999hq,Denner:2002ii,Duplancic:2003tv,Giele:2004iy,Ossola:2006us}.
We refer to the integrals with external momenta in a generic $d$-dimensional space as conventional dimensional regularization (CDR) integrals, while to those constrained to $d_0=4$ external dimensions as `t Hooft-Veltman (tHV) integrals.
Recently, this constraint has been applied at two loops in Ref.~\cite{Bargiela:2024rul}.
Explicitly, any integrals with more than 11 propagators can be reduced to 12 top sector topologies and their subsectors.
This puts an upper bound on the structure of all two-loop Feynman integrals in the Standard Model, which we will investigate in this work.
Moreover, for integrals with at least 9 denominators, the tHV decomposition improves the performance of the associated IBP reduction, see Ref.~\cite{Bargiela:2024rul}.
Here, we will show that the tHV decomposition also reduces the complexity of the numerical evaluation of Feynman integrals.

Firstly, we will explicitly present the functionally distinct finite basis for all two-loop Feynman integrals.
It is well-known that, at one loop, there are 5 finite basis integrals, i.e. tadpole, bubble, triangle, box, and pentagon~\cite{Melrose:1965kb,Passarino:1978jh,vanNeerven:1983vr,Bern:1992em,Binoth:1999sp,Fleischer:1999hq,Denner:2002ii,Duplancic:2003tv,Giele:2004iy,Ossola:2006us}.
At two loops, there has been a recent major progress in bounding the spectrum of independent Feynman integrals.
The planar top sectors of the finite basis have been found in Ref.~\cite{Gluza:2010ws}, while the nonplanar in Ref.~\cite{Kleiss:2012yv}.
In addition, for lower multiplicity, additional reductions have been introduced e.g. in Refs~\cite{Mastrolia:2011pr,Mastrolia:2013kca}.
The explicit reduction from CDR to tHV integrals has been given in Ref.~\cite{Bargiela:2024rul}.

Secondly, we will indicate that our two-loop finite basis can be further constrained by considering only the integrals which can contribute to the finite part of any two-loop amplitude.
It is well-known that all one-loop integrals beyond four-point can be expressed as a linear combination of box and it's subsector integrals, up to $\mathcal{O}(\epsilon)$ corrections~\cite{Bern:1993kr}, i.e. they are evanescent.
At two loops, a similar constraint has been investigated in Ref.~\cite{Feng:2012bm}.
In general, evanescent integrals has been systematically studied in Ref.~\cite{Gambuti:2023eqh}.

Thirdly, we will analyze the spectrum of special functions and the corresponding geometries appearing in any two-loop scattering amplitude in $d=4-2\ep$.
It is well-known that, at one-loop, this spectrum consists solely of Multiple Polylogarithms~\cite{Chen:1977oja,Goncharov:1995ifj}, among which only logarithm and di-logarithm contribute to the finite part of any one-loop amplitude~\cite{Ellis:2007qk}.
In the Standard Model calculations at two-loop order, the spectrum is usually obtained for each physical process separately by explicitly solving all the independent integrals, see an example review in Ref.~\cite{Heinrich:2020ybq}.
In more symmetric theories the knowledge about such a spectrum is much wider.
For example, the analysis of the full iterated integral structure has been recently completed in planar $\mathcal{N}=4$ super-Yang-Mills theory in $d=4$ in Ref.~\cite{Spiering:2024sea}.
In addition, the geometries stemming from Feynman integrals contributing to the classical dynamics of a black-hole two-body system in the post-Minkowskian expansion of general relativity have been recently investigated to three loops in Ref.~\cite{Frellesvig:2024zph}.

The remainder of this paper is organized as follows.
In Sec.~\ref{sec:fin}, we present the functionally distinct finite basis for all two-loop Feynman integrals.
In Sec.~\ref{sec:evan}, we indicate that all two-loop Master Integrals, expressed in an appropriate basis, with more than 8 denominators do not contribute to the finite part of any two-loop amplitude.
In Sec.~\ref{sec:num}, we elaborate on the application of the tHV decomposition to numerically evaluate Feynman integrals.
In Sec.~\ref{sec:geom}, we analyze the spectrum of special functions and the corresponding geometries appearing at two-loops.
In Sec.~\ref{sec:concl}, we summarize our results, as well as we suggest potential future directions and applications.

\section{Finite basis}
\label{sec:fin}

In this section, we present a finite basis of Feynman integrals with arbitrary kinematics which spans the whole transcendental space of all infinitely-many two-loop Feynman integrals in $d=4-2\ep$.

Consider a generic CDR Feynman integral
\begin{equation}
    \intl_{\vec{\nu},\text{CDR}} 
    = \int \frac{d^d k_1}{(2\pi)^d} \, \frac{d^d k_2}{(2\pi)^d} \, \intd_{\vec{\nu},\text{CDR}} \,,
    \label{eq:intlDef}
\end{equation}
with the integrand
\begin{equation}
    \intd_{\vec{\nu},\text{CDR}} = \prod_{i=1}^{S_\text{CDR}} \mathcal{D}_i^{-\nu_i} \,,
    \label{eq:intdDef}
\end{equation}
where $\nu_i$ are integer powers and all the $S_\text{CDR}$ generalized propagators are quadratic, i.e. $\mathcal{D}_i = q_i^2 - m_i^2$, with masses $m_i$ and momenta $q_i$ which depend on external momenta $p_n$ and loop momenta $k_l$, both treated as $d$-dimensional.
Following Ref.~\cite{Bargiela:2024rul}, when treating all the external momenta $p_n$ as purely four-dimensional, any CDR integrand $\intd_{\vec{\nu},\text{CDR}}$ can be linearly decomposed into a sum of tHV Feynman integrands $\intd_{\vec{\mu}}$, i.e.
\begin{equation}
    \intd_{\vec{\nu},\text{CDR}} = \sum_{\vec{\mu}} c_{\vec{\mu}} \, \intd_{\vec{\mu}} \,,
    \label{eq:PF2L}
\end{equation}
where the coefficients $c_{\vec{\mu}}$ are rational functions of external kinematics.
Indeed, it is possible to form at most $S=11$ generalized propagators from scalar products which involve exactly 2 loop momenta $k_1,k_2$ and at most 4 independent external momenta $p_1,p_2,p_3,p_4$.
All further propagators are not linearly independent.
Note that the $z$ variables introduced in Ref.~\cite{Bargiela:2024rul} in the tHV propagators are constructed as in the following example $k_1-k_2+p_1+p_2+p_3+p_4+p_5+p_6 = k_1-k_2+\sum_{i=1}^{4} z_i \, p_i$.
It has been shown in Ref.~\cite{Bargiela:2024rul} that there are exactly 12 diagrammatically distinct top sector topologies with 11 propagators, referred to as the \textit{finite basis topologies}, as depicted in the last 12 rows of Tab.~\ref{tab:MIs}.
In comparison, at one-loop, there is only one finite basis topology, i.e. the pentagon.
Since, for the tHV integrals, an upper bound on their structure is clear, throughout this work, we will only analyze the tHV integrals, and not the CDR ones.

In this work, we are after finding a finite basis of Feynman integrals  that spans the space of all two-loop Feynman integrals, similarly as how the five finite basis integrals, i.e. the tadpole, bubble, triangle, box, and pentagon, span the one-loop integral space~\cite{Melrose:1965kb,Passarino:1978jh,vanNeerven:1983vr,Bern:1992em,Binoth:1999sp,Fleischer:1999hq,Denner:2002ii,Duplancic:2003tv,Giele:2004iy,Ossola:2006us}. 
We consider here only the all-off-shell Feynman integrals with pairwise distinct values of the squared external momenta, since we assume the standard property of integral topologies that no new Master Integrals appear for more degenerate kinematics.
Also, without loss of generality, we focus on integrals that cannot be related to each other by permutations of the external momenta. 
Other integrals can either be obtained from the resulting basis integrals by permuting the external momenta, or they are functionally simpler due to the degeneracy of the external kinematics. 
These basis integrals can be extracted by analyzing the diagrammatically distinct Feynman integral graphs. At two loops, there are only 84 such graphs, as depicted in Tab.~\ref{tab:MIs}, which are all subgraphs of the 12 top sector graphs with 11 propagators. Note that only 80 graphs out of these 84 have a distinct set of generalized propagators.
Indeed, the sunrise with only one external leg, and vacuum graphs have the same set of propagators. 
In addition, at two loops, Feynman diagrams with one squared propagator, the so called dotted propagator, are allowed, i.e. the one-loop two-, three-, and four-point graphs with a one-loop bubble correction on an internal line.

To explicitly find the finite basis integrals, we first consider all integrals in the top sector defined by each of the 80 graphs. This is equivalent to requiring the powers $\mu_i$ of generalized propagators $\mathcal{D}_i$ appearing in a graph to be strictly positive, while those of all other propagators, i.e. the Irreducible Scalar Products (ISPs), to be non-positive. In order to find a linearly independent set of Master Integrals (MIs) in each such top sector, we use the Integration-By-Parts (IBP) identities~\cite{Tkachov:1981wb,Chetyrkin:1981qh,Laporta:2000dsw}
\begin{equation}
    \int \frac{d^d k_1}{(2\pi)^d} \, \frac{d^d k_2}{(2\pi)^d} \, \frac{\partial}{\partial k_l^\sigma} \, (q^\sigma \intd_{\vec{\mu}}) \, = 0 \,,
\end{equation}
with $l \in \{1,2\}$ and $q \in \{k_1,k_2,p_1,p_2,p_3,p_4\}$.
We perform the IBP reduction on a maximal cut associated with each graph.
Note that before explicitly evaluating the integrals, the concept of a cut propagator serves only as a label.
To identify a set of MIs, an IBP reduction on a numerical kinematic probe over one finite field~\cite{vonManteuffel:2014ixa,Peraro:2016wsq} is sufficient. 
By running \texttt{FIRE6}~\cite{Smirnov:2019qkx,Smirnov:2023yhb} and \texttt{Kira}~\cite{Maierhofer:2017gsa,Klappert:2020nbg,Klappert:2019emp} on each of the 80 top sectors for all off-shell legs, we obtained a total of 347 MIs for internal lines with arbitrary masses and 274 MIs for fully massless internal lines. 
The MIs are chosen to favor numerators over dots. And the numerators of these MIs are listed in Tab.~\ref{tab:MIs}. 
We also provide computer-readable expressions for both sets, along with the corresponding tHV generalized propagator definitions, in the ancillary files.
We confirmed that there are no extra MIs by running the IBP reductions multiple times by gradually increasing the degree of ISPs. 

We argue that the identified sets of MIs represent the basis integrals we are seeking, which is supported by the maximal cut reduction technique and a recently proposed transverse integral decomposition related to integral reductions in Ref.~\cite{Chestnov:2024mnw}. The maximal cut reduction technique refers to the idea that the MIs for a top sector and its subsectors can be efficiently identified by performing IBP reductions on the maximal cut of each sector. This approach was first introduced and implemented in Ref.~\cite{Georgoudis:2016wff} and can be easily applied using any IBP reduction tool. The transverse integral decomposition can be illustrated with an explicit example. Specifically, we consider the sunrise graph on its maximal cut to identify all MIs in its top sector. Since the sunrise graph is a subsector of, for example, the double-box graph without any cut constraint, there are numerators in the double-box topology that belong to a transverse space relative to the single independent external momentum in the sunrise graph. It has been shown in Ref.~\cite{Chestnov:2024mnw} that integrals with such transverse numerators are always reducible and therefore do not generate any new MIs, modulo crossings, with respect to the MIs from the sunrise graph.

Therefore, the set of 347 MIs is the functional finite basis of all the two-loop Feynman integrals with quadratic propagators in $d=4-2\ep$. This means that it spans the space of all the transcendental functions, without specifying a parametrization.
As such, our finite MI basis is an upper bound on the set of functionally distinct MIs for a specific physical process.
Indeed, the complete set of MIs for a fixed kinematics will be larger by accounting for all the crossings, and it may involve simpler transcendental functions due to degenerate kinematics.
Note that, in principle, if all of the functional finite basis integrals became available in their full-$\ep$ analytic form, then all their degenerate limits could be extracted.
We also point out that explicitly relating an arbitrary integral to our finite MI basis still requires an IBP reduction without imposing any cut constraints, which remains a bottleneck. 

Let us briefly comment on the properties of the obtained finite MI basis.
Per each relative graph top sectors, there are at most 17 MIs in both fully internally massive and massless cases.
The number of sectors with exactly 0 MIs increases from 0 to 11 when putting all the internal masses to 0.
The number of sectors with exactly 1 MI decreases from 37 to 31 under the same condition.
In total, the number of MIs decreases in 32 sectors when considering only massless propagators.
Per each sector, the number of ISPs ranges from 0 to 5.
In Tab.~\ref{tab:MIs}, we present an example choice of the MI basis in each sector.
It only consists of ISPs, without any dots, and the numerators are arranged starting from the lowest-rank monomials in the available ISP variables.

\section{Evanescent sectors}
\label{sec:evan}

In this section, we show on a numerical probe that all the two-loop Master Integrals, expressed in an appropriate basis, with more than 8 denominators do not contribute to the finite part of any two-loop amplitude. 

It is well-known that all one-loop integrals beyond four-point can be expressed as a linear combination of box and it's subsector integrals, up to $\mathcal{O}(\epsilon)$ corrections~\cite{Bern:1993kr}.
It can be proven in the following steps.
Firstly, all the CDR integrals beyond five points are reducible in tHV to at most five-point integrals.
Secondly, all the five-point integrals with nontrivial numerators can be decomposed using Passarino-Veltman reduction~\cite{Passarino:1978jh} into the pentagon integral and its subsectors, both with $\ep$-independent coefficients.
Thirdly, the single MI in the pentagon sector can be chosen to be evanescent, i.e. to have vanishing poles and finite part, such that it starts contributing at $\mathcal{O}(\ep)$.
Since there are no poles in $\ep$ explicitly generated by Feynman rules and Dirac algebra, then there is no contribution from the pentagon sector to the finite part of any one-loop amplitude.

In what follows, we will apply a similar reasoning at two loops.
We start by reiterating that any higher-point two-loop integrals can be decomposed into subsectors of the 12 two-loop finite basis topologies~\cite{Bargiela:2024rul}, as depicted in the last 12 rows of Tab.~\ref{tab:MIs}.
We will analyze all their subsectors with more than 8 denominators, i.e. 9, 10, and 11.
We expect all of them to be evanescent since there are only $4L$ linearly independent propagators at $L$ loops in exactly $d=4$ dimensions.
As such, this effect starts appearing for integrals with the number of denominators $n+3(L-1)$ exceeding $4L$, i.e. with at least $n=L+4$ legs.
We start by running the IBP reduction using \texttt{Kira}~\cite{Maierhofer:2017gsa,Klappert:2020nbg,Klappert:2019emp} for each of the 12 finite basis topologies while dropping subsectors with 8 or less denominators. 
To make the reduction feasible while preserving the representative properties of the IBP system, we select a generic numerical probe for the external kinematics and focus on all massless propagators. Specifically, we consider the four basis momenta $p_1,p_2,p_3,p_4$ on-shell and the other legs $p_{i>4}$ off-shell, and retain the full parametric dependence on $d$. 
Note that the integrals in this kinematics are in general singular in $\ep$.
In addition, we point out that we analyze here also these subsectors which may have the same graph but differ in kinematic parametrization.
We obtained $\{{167, 167, 167, 295, 285, 287, 332, 327, 332, 445, 429, 445}\}$ MIs without any dots for the last 12 topologies in Tab.~\ref{tab:MIs}, respectively, while still dropping subsectors with 8 or less denominators.

The obtained Laporta basis of MIs does not explicitly evaluate to evanescent expressions.
Indeed, we have to choose a basis which is locally IR and UV finite, and then show evanescence.
Let us start with the IR singularity analysis.
It is convenient to note that there are no IR singularities in $d+2=6-2\ep$ dimensions, provided that the integrals do not have dots~\cite{Dyson:1949ha,Weinberg:1959nj}.
This caveat is not a problem for our purposes since all the graphs with a one-loop bubble correction to an internal line are all linearly reducible to less than 9-denominator sectors, see Ref.~\cite{Bargiela:2024rul}.
Therefore, for each finite basis topology, we seek a basis of finite MIs $M^{(\text{6D})}_j$ that satisfies
\begin{equation}
    M_i^{(\text{4D})} = \ep \, A^{\text{(fin)}}_{ij} M^{(\text{6D})}_j + \text{subsectors} \,,
\end{equation}
where the matrix of coefficients $A^{\text{(fin)}}_{ij}$ is regular in $\ep$, and we drop all the subsectors with 8 or less denominators.
As a candidate MI basis $M^{(\text{6D})}_j$, we simply chose the same Laporta MIs as obtained from \texttt{Kira} but shifted to $6-2\ep$ dimensions.
We found the dimensional shift relations~\cite{Tarasov:1996br,Baikov:1996iu} between our MIs using \texttt{LiteRed}~\cite{Lee:2013mka}.
Then, we IBP reduced the resulting integrals into our $M^{(\text{6D})}_j$ MI basis using \texttt{Kira}.
As a result, we indeed observed that $\ep \, A^{\text{(fin)}}_{ij}$ vanishes when $\ep \to 0$.
We stress that in order to find a MI basis with evanescent coefficients it was enough to just shift the Laporta MIs into six dimensions.
It would be interesting to see if this property holds analytically for any choice of tHV MIs.

Having finished the IR analysis, we still have to guarantee the UV finiteness.
This can be achieved by demanding for our MIs $M^{(\text{6D})}_j$ to be power-counting UV finite in both internal loops separately, as well as all together.
In our case, the only source of UV singularities could be higher-degree numerators.
It can be shown that $M^{(\text{6D})}_j$ are also UV finite if the original Laporta basis numerators are all linearized according to $(k_l+q)^2 \to k_l \cdot q$ and $k_l^2 \to c_{l00} \, k_1 \cdot k_2 + \sum_{j \in \{1,2\},i \in \{1,2,3,4\}} c_{lji} \, k_j \cdot p_i$, where $c_{lji}$ are any coprime numerical coefficients, and $l \in \{1,2\}$.
Indeed, with such a locally finite MI basis $M^{(\text{6D})}_j$ choice, the coefficients $\ep \, A^{\text{(fin)}}_{ij}$ vanish when $\ep \to 0$.

\begin{figure}[!h]
    \centering
    \includegraphics[width=0.99\textwidth]{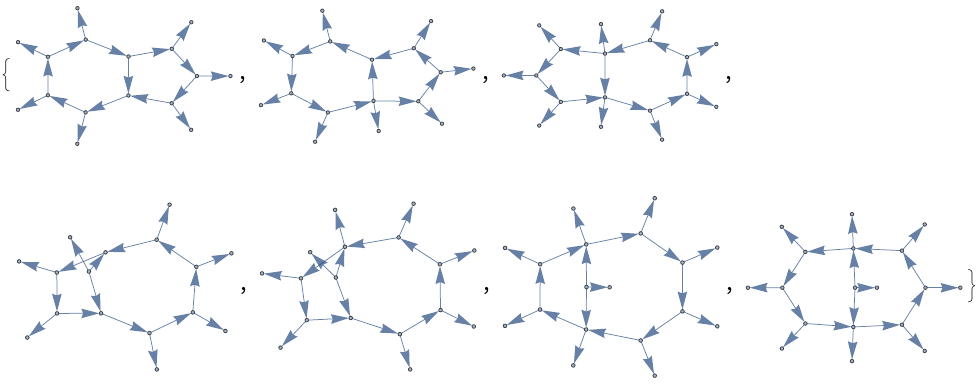}
    \caption{Doubly-evanescent sectors found on a numerical probe.}
    \label{fig:evan2}
\end{figure}

This concludes the construction of an evanescent MI basis for sectors with more than 8 denominators, on a numerical probe.
Interestingly, we found that, when considering the IBP reduction in each sector separately, there are 7 sectors, as depicted in Fig.~\ref{fig:evan2}, each with 10 denominators, which are doubly evanescent, i.e. $M_i^{(\text{4D})} = \ep^2 \, A^{\text{(fin)}}_{ij} M^{(\text{6D})}_j + \text{subsectors}$.
We point out that, since our analysis was performed on a numerical kinematic sample, it is not a formal proof.
Indeed, it would require solving the IBP system analytically for all the kinematic parameters in every possible degenerate limit separately, which is prohibitive at the moment.
Recently, for specific topologies and kinematics, the evanescence has been shown using other methods in Ref.~\cite{Abreu:2024fei} for fully massless on-shell six-point double-pentagon and hexa-box CDR sectors.
It would be interesting to confirm our numerical observation fully analytically, including the existence of the doubly-evanescent sectors.

Finally, we connect the discussion of MIs to scattering amplitudes.
We expect the finite part of any two-loop amplitude to depend only on MIs with at most 8 denominators, i.e.
\begin{equation}
\begin{split}
   \mathcal{A}^{(2)} 
    &= a_i \, M^{(2)}_{\le 8 \text{den},i} + b_j \, M^{(2)}_{> 8 \text{den}, \text{evan},j} \\
    &= a_i \, M^{(2)}_{\le 8 \text{den},i} + \mathcal{O}(\ep)\,,
\end{split}
\end{equation}
where $a_i$ and $b_j$ are rational functions of kinematic invariants.
They arise from the amplitude integrand in a finite manner, and from the IBP reduction which may induce poles in $\ep$.
Thus, we have to argue that $b_j$ are finite in $\ep$.
Similarly as in Ref.~\cite{Abreu:2024fei}, this can be shown by considering all the possible numerators allowed in renormalizable theories.
The relevant maximal total numerator degree is bounded by the occurrence of momenta in numerators of Feynman rules applied to the particle dressings of the two-loop finite basis topologies.
By IBP reducing all of the allowed integrals using \texttt{Kira} onto the evanescent MIs basis constructed above, we observed that the IBP coefficients are finite in $\ep$.
Beyond 11 denominators, for each additional denominator, one more numerator power is allowed.
Importantly, due to the linear dependence of the additional denominator on the basis 11 denominators, the one additional numerator can be always canceled, e.g.
\begin{equation}
\begin{split}
   \frac{1}{\mathcal{D}_1 \cdots \mathcal{D}_{11}}
   \frac{\mathcal{N}}{\mathcal{D}_{12}}
   = \frac{1}{\mathcal{D}_{12}}
   \frac{a_0 + \sum_{i=1}^{11} a_i \mathcal{D}_i}{\mathcal{D}_1 \cdots \mathcal{D}_{11}}
   = \frac{1}{\mathcal{D}_{12}} \bigg( \frac{a_0}{\mathcal{D}_1 \cdots \mathcal{D}_{11}}
   + \sum_{i=1}^{11} \frac{a_i}{\mathcal{D}_1 \cdots \hat{\mathcal{D}}_i \cdots \mathcal{D}_{11}} \bigg) \,,
\end{split}
\end{equation}
where $\hat{\mathcal{D}}_i$ denotes an absence of the denominator $\mathcal{D}_i$.
We point out that this numerator reduction is also important for decreasing the associated computational complexity.
Finally, the remaining linearly dependent denominators can be partial fractioned into the subsectors of the finite basis topologies, according to Ref.~\cite{Bargiela:2024rul}.
This concludes the argument that all the two-loop Master Integrals, expressed in an appropriate basis, with more than 8 denominators, i.e. 9, 10, or 11, would not contribute to the finite part of any two-loop amplitude.

\section{Improving the numerical evaluation}
\label{sec:num}

In this section, we elaborate on the application of the tHV decomposition to numerically evaluate Feynman integrals.
Following Ref.~\cite{Bargiela:2024rul}, we show examples of removing redundant numerators at two-loop order to improve the efficiency of \texttt{AMFlow}~\cite{Liu:2022chg}, and removing redundant denominators at one loop to improve the efficiency of the \texttt{DCT}~\cite{Huang:2024qan} package.

\subsection{tHV for AMFlow}

\begin{figure}[!h]
    \centering
    \includegraphics[width=0.2\textwidth]{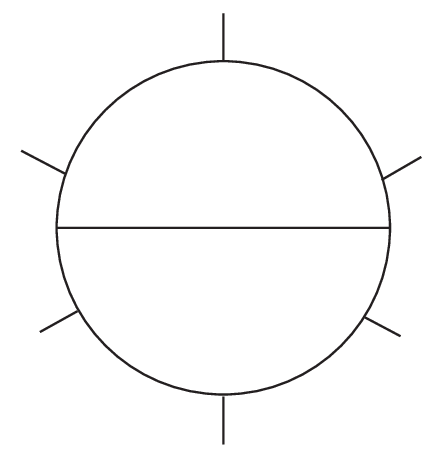}
    \caption{Two-loop six-point topology with a few off-shell legs evaluated with \texttt{AMFlow}.}
    \label{fig:2L6p}
\end{figure}

\noindent
In Ref.~\cite{Bargiela:2024rul}, it has been exemplified that using tHV instead of CDR parametrization of Feynman integrals reduces the complexity of the associated IBP reduction.
Indeed, at two loops, this is caused by the drop in the number of numerators, starting from 9-denominator topologies, as well as denominators, starting from 12-denominator topologies, in tHV comparing to CDR.
This property can be applied to simplify also the numerical evaluation of integrals in approaches based on the IBP reduction, such as \texttt{AMFlow}~\cite{Liu:2022chg}.

This idea has been already used in numerical checks of the analytic computation of the fully on-shell massless two-loop six-point double-pentagon MIs in Ref.~\cite{Henn:2025xrc}, as depicted in Fig.~\ref{fig:2L6p}.
As an example, we consider here the same topology but with two off-shell legs. On a kinematic sample with one-digit integers, we numerically evaluated the top sector integrals using \texttt{AMFlow} with the \texttt{Kira} backend to 20 digits at the first 5 leading orders in $\epsilon$, utilizing 16 CPU cores over the course of approximately 30 hours.
In comparison, running \texttt{AMFlow} on the same number of cores but without the tHV decomposition did not even complete the IBP reduction after 1 week of evaluation time.

In addition, we used this procedure to numerically verify if the MI basis in the double-pentagon sector chosen in Sec.~\ref{sec:evan} is really finite.
As an example, we evaluated all the 5 top sector MIs $M^{(\text{6D})}_j$ in this topology with one off-shell leg and internally-massless kinematics
\begin{align}
    M&^{(\text{6D})}_j
    \in \int \frac{d^{d+2} k_1}{(2\pi)^{d+2}} \frac{d^{d+2} k_2}{(2\pi)^{d+2}} \nonumber \\
    & \times \,
    \frac{\{1, k_1 \cdot p_4, k_2 \cdot p_1, (k_1 \cdot p_4)^2, k_1 \cdot p_4 \, k_2 \cdot p_1\}}{k_1^2 k_2^2 (k_1-k_2)^2 (k_1+p_1)^2 (k_1+p_{12})^2 (k_1+p_{123})^2 (k_2+p_{123})^2 (k_2+p_{1234})^2 (k_2+z_i p_i)^2} \,,
\end{align}
where $i \in \{1,2,3,4\}$.
On a kinematic sample with three-digit rationals, we evaluated these MIs numerically using \texttt{AMFlow} with \texttt{Kira} backend, utilizing 16 CPU cores for about 100 hours.
As a result, we verified that all their poles in $\ep$ cancel to 20 digits, and the first nonvanishing order is $\mathcal{O}(\ep^0)$.

\subsection{Partial fraction for DCT}

\begin{figure}[!h]
    \centering
    \includegraphics[width=0.2\textwidth]{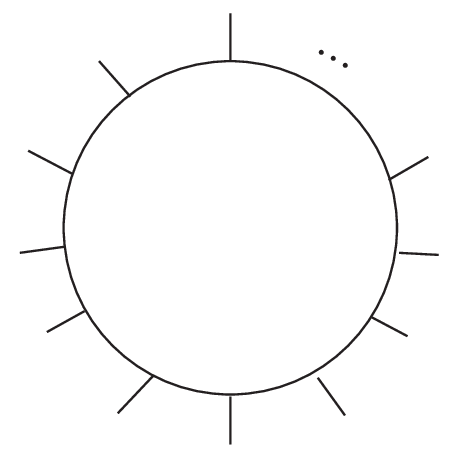}
    \caption{One-loop high-multiplicity topologies.}
    \label{fig:1LMp}
\end{figure}

\begin{figure}[!h]
    \centering
    \includegraphics[scale=0.7]{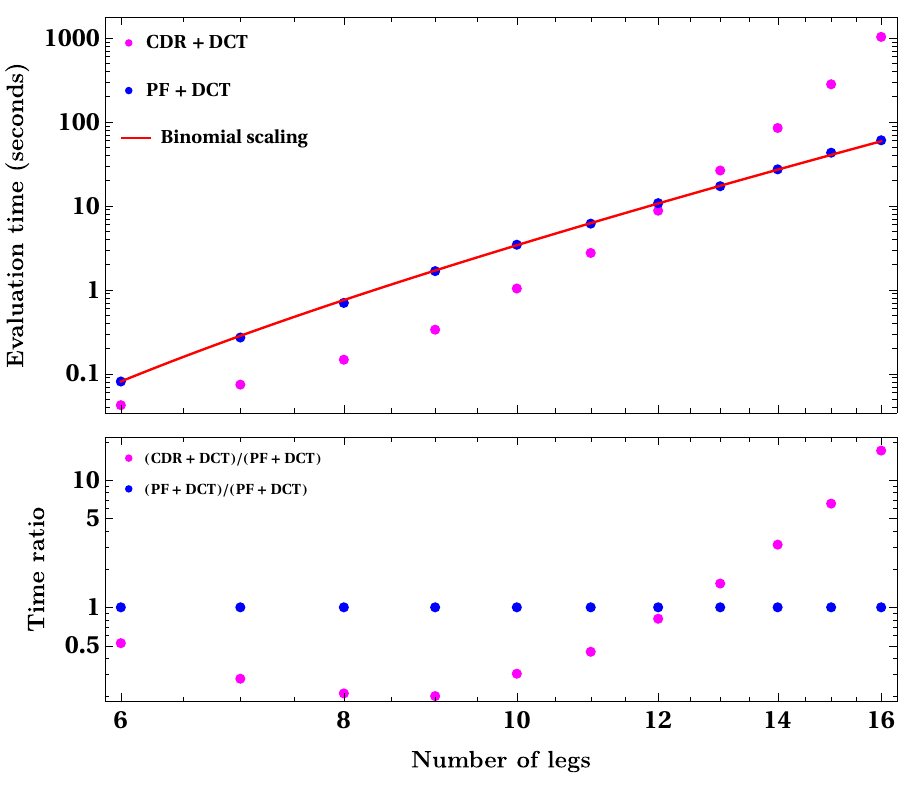}
    \caption{Comparison of evaluation efficiency as a function of the number of legs for the "CDR + DCT" and "PF + DCT" methods. "Binomial scaling" stands for the binomial formula in Eq.~\eqref{eq:binomial_1L} normalized to the evaluation time for $n=6$ of "PF + DCT" method. } 
    \label{fig:DCT}
\end{figure}

\noindent
Recently, a method called the dimension-changing transformation (DCT)~\cite{Huang:2024qan} has emerged as an effective approach for the efficient numerical evaluation of one-loop multi-point integrals to higher orders in $\epsilon$. In this section, we show that combining the closed partial fraction form proposed by us in Ref.~\cite{Bargiela:2024rul} with the DCT method — which we refer to as "PF + DCT" — can further enhance the efficiency of numerical evaluation compared to evaluating CDR integrals using the DCT method, referred to as "CDR + DCT".

As shown explicitly in Ref.~\cite{Bargiela:2024rul}, a $L$-loop CDR integral with $n+ 3(L-1)$ denominators and $\frac{1}{2} (L-1) (L+2 n-6)$ ISPs can be decomposed as $\binom{n+ 3(L-1)}{4 L + L(L+1)/2}$ integrals with $4 L + L(L+1)/2$ propagators, through a closed partial fraction form. For a one-loop $n$-point integral with $n$ propagators, as shown in Fig.~\ref{fig:1LMp}, the aforementioned binomial reads
\begin{equation}
\label{eq:binomial_1L}
    \binom{n}{5} =\frac{1}{120} (n-4) (n-3) (n-2) (n-1) n \,.
\end{equation}
In order to compare "CDR + DCT" and "PF + DCT", we numerically evaluate the single top sector one-loop all-off-shell and internally-massless integral with all propagator powers being 1 for $6,7, 8, \cdots, 16 $ legs using both methods. For "PF + DCT," we sequentially evaluate multiple one-loop pentagon integrals obtained from partial fraction decomposition. Thanks to the closed partial fraction formula in Ref.~\cite{Bargiela:2024rul}, the time required for partial fraction decomposition is negligible. For instance, the partial fraction decomposition with analytic coefficients for the CDR integral with 16 legs takes only 12 seconds on a standard laptop. 

The comparison between the results is shown in Fig.~\ref{fig:DCT}. The upper panel presents the evaluation time for a single numerical sample, measured in seconds, as a function of the number of legs for the two methods. The lower panel shows the time ratio relative to the evaluation time of the "PF + DCT" method. We checked explicitly that both these methods give consistent numerical results. In terms of an absolute evaluation time, for $n > 12$, the "PF + DCT" works better than "CDR + DCT". As $n$ increases, the scaling behavior of "CDR + DCT" is much worse than "PF + DCT", for example the time ratio is 1.53 for $n=13$, while it becomes 17 for $n=16$. For $n\leq 12$, the "CDR + DCT" method is quite fast, and it evaluates in less than 10 seconds for one numerical sample. While evaluating one integral in "PF + DCT" is much faster, evaluating 792 such integrals for $n=12$ sequentially is a bit longer than "CDR + DCT". For $n=9$, "CDR + DCT" is five times faster than "PF + DCT". Nevertheless, we still recommend using "PF + DCT" for $n \leq 12$ since evaluating a one-loop pentagon integral multiple times is more stable, requires less memory, and allows for a massively parallel evaluation. The red line in Fig.~\ref{fig:DCT} is "Binomial scaling", which represents the scaling behavior of formula~\eqref{eq:binomial_1L} normalized to evaluation time of "PF + DCT" for $n=6$, i.e.,
\begin{equation}
    T_n = \frac{\binom{n}{5}}{\binom{6}{5}} T_6 \,.
\end{equation}
The "Binomial scaling" agrees well with the explicit results in "PF + DCT" (blue dots in Fig.~\ref{fig:DCT}), this allows us to easily predict the evaluation time for integrals with any number of legs in "PF + DCT", for example, $T_{20} =208 \text{ seconds} $ and $T_{30} = 1913 \text{ seconds}$. 

Recently, a new approach for numerically evaluating integrals beyond one-loop was proposed in Ref.~\cite{Huang:2024nij}, with further developments along this direction presented in Ref.~\cite{Chen:2025lfi}. By combining these methods with the closed formula for partial fraction decomposition in the tHV scheme, it is expected that the efficiency of numerical evaluation for beyond one-loop integrals can be further enhanced, similar to the case of the "PF + DCT" method for one-loop integrals discussed above. However, it is beyond the scope of this paper and we leave it for future studies.

\section{Special functions and geometries}
\label{sec:geom}

In this section, we analyze the spectrum of special functions and the corresponding geometries appearing in any two-loop scattering amplitude in $d=4-2\ep$.

Scattering amplitudes can be written as a linear combination of transcendental MIs with rational function coefficients.
Since the latter is well-understood, we will focus here on the iterated integral structure of the former.
It is well-known that the spectrum of special functions appearing in the solution of any given Feynman integral arises from all of its subsectors considered on their relative maximal cuts.
As discussed in Sec.~\ref{sec:fin}, there are exactly 80 graph sectors at two loops in $d=4-2\ep$ which lead to functionally distinct expressions.

It is convenient to consider the required relative maximal cuts of MIs in these sectors in the Baikov representation~\cite{Baikov:1996cd}.
For a given integral topology with $n$ independent external momenta, we have
\begin{equation}
    \intl_{\vec{\mu},\text{max}} 
    = \int \frac{d^d k_1}{(2\pi)^d} \, \frac{d^d k_2}{(2\pi)^d} \prod_{i=1}^{S_{\text{den}}} \delta(\mathcal{D}_i) \prod_{j=S_{\text{den}}+1}^{S_{\text{den}}+S_{\text{ISP}}} \mathcal{D}_j^{-\mu_j} 
    \sim \int \mathcal{B}(\vec{x})^\gamma \prod_{j=1}^{S_{\text{ISP}}} x_j^{-\mu_j} d x_j \,,
    \label{eq:Baikov}
\end{equation}
where we dropped all factors which would not contribute to the Baikov integral, and the Baikov variables are the propagators $x_i=\mathcal{D}_i=q_i^2-m_i^2$, while $\vec{x}= (x_1,x_2,\cdots, x_{S_{\text{ISP}}})$ are the unintegrated Baikov variables corresponding to ISPs. And $S_{\text{den}}$ is the number of denominators, $S_{\text{ISP}}$ is the number of ISPs, $\vec{\mu}$ has $S_{\text{den}}$ first indices equal to one, and the remaining $S_{\text{ISP}}$ are nonpositive, $\mathcal{B}$ is the Baikov polynomial evaluated at the maximal cut, and $2\gamma=d-n-3$.
We derive all of the relevant Baikov polynomials using an in-house routine.
Alternatively, it is often beneficial to use the loop-by-loop (LBL) Baikov representation~\cite{Frellesvig:2017aai}.
Using the LBL \texttt{BaikovPackage} package~\cite{Frellesvig:2024ymq}, we obtained LBL Baikov polynomials for some topologies, and for the remaining ones, we used the Baikov representation.
In both approaches, we found that, independently of the values of internal masses of the propagators, the maximal total degree among all of the Baikov polynomials is 4, while the maximal number of Baikov variables, i.e. ISPs, is 5, see a summary in Tab.~\ref{tab:MIs}. An example for the obtained Baikov polynomial with total degree 4 reads $\mathcal{B}(\vec{x}) = c_1 x_1^4 + c_2 x_1^3 x_2 + c_2 x_1^2 x_2^2 + c_3 x_1 x_2 x_3 x_4 + c_4 x_2^2 + c_5 + \dots$, where $c_i$ are coefficients dependent only on the external kinematics.

Having found the Baikov polynomials, let us investigate the iterated integral structure of the Baikov integral.
It is well-known that the spectrum of special functions stemming from the Baikov integral does not depend on the integration contour.
In addition, due to the dimensional shift relations, such spectrum also does not depend on the power $\gamma$ of the Baikov polynomial.
Moreover, for phenomenological applications, we are interested only in the $\ep$-expanded form of Feynman integrals.
Therefore, we can choose the simplest and representative contour and power, e.g. a closed contour in one of the variables $w \in \vec{x}$, and $\gamma=-1$.
Using Cauchy's residue theorem, the Baikov integral can be integrated in one of the variables $w$ by summing over the residues, which results in a function $f$ of $\vec{y} = \vec{x} \setminus \{w\}$, i.e.
\begin{equation}
    \oint \frac{dw}{\mathcal{B}(w,\vec{y})} 
    = 2 \pi i \sum_{w^* \, : \, \mathcal{B}(w^*,\vec{y}) = 0} \text{Res}_{w \to w^*} \frac{1}{\mathcal{B}(w,\vec{y})}
    = f(\vec{y}) \,.
    \label{eq:Res}
\end{equation}
Since all of our Baikov polynomials in any of the variables separately have degree either 2 or 4, then the function $f$ can functionally involve, in addition to rational terms, square roots of a polynomial $\sqrt{P(\vec{y})}$, or iterated roots, respectively.
For Baikov polynomial, we computed the residues in each variable separately and chose the one which decreases firstly the number of iterated roots, and secondly the degree of the polynomial $P$ under the square root in the result $f$.
Note that we do not keep any MI numerators in contour integral above.
We can drop these terms because these numerators cannot further complicate the roots in functions $f$.

\begin{figure}[!h]
    \centering
    \includegraphics[width=0.99\textwidth]{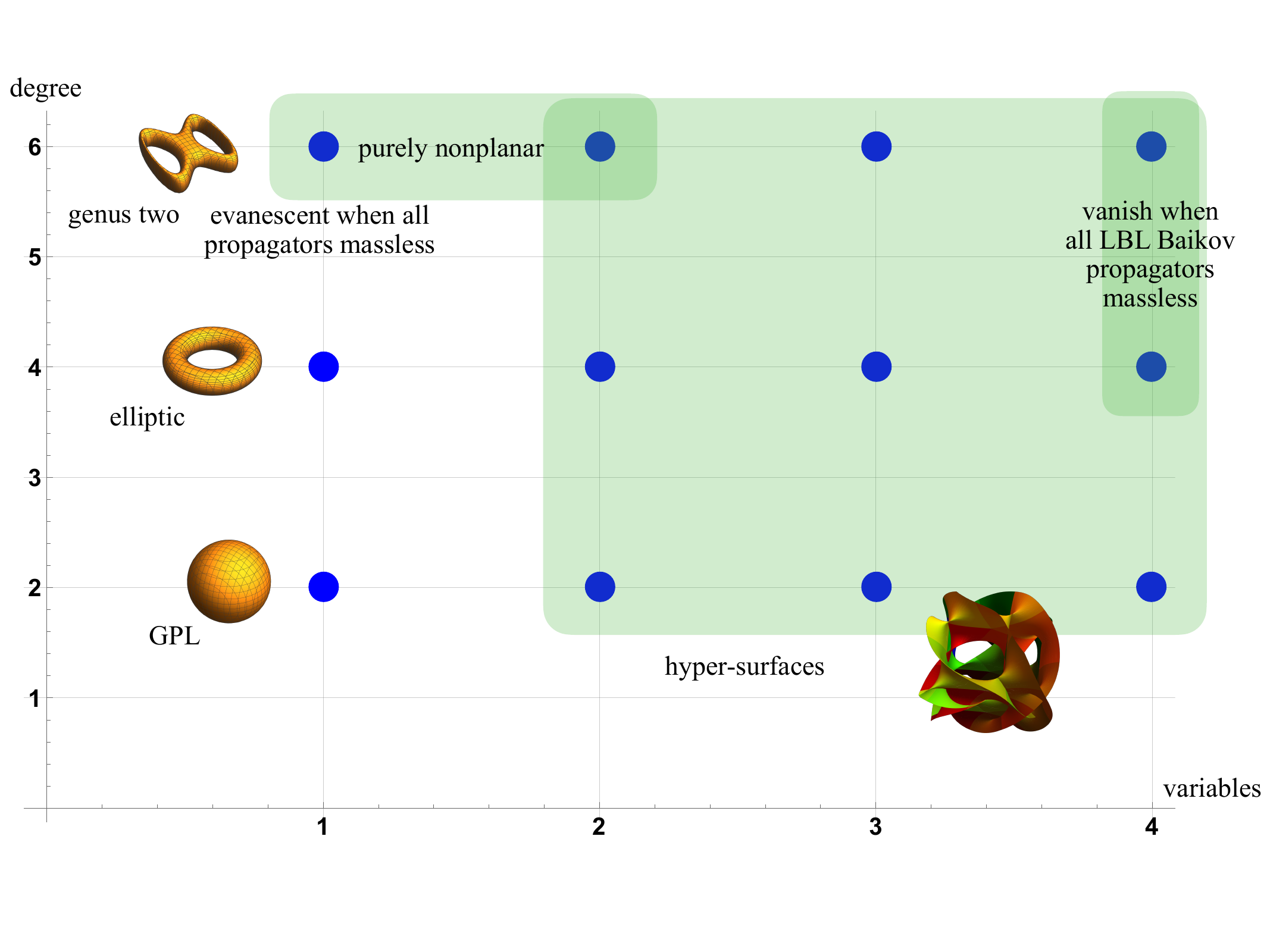}
    \caption{The number of variables and the total degree of polynomials encountered under square roots after taking one Residue in Baikov representation at the maximal cut for all the two-loop sectors, together with the corresponding geometries.}
    \label{fig:geoms}
\end{figure}

Let us start by analyzing the sectors which lead only to square roots $\sqrt{P(\vec{y})}$ without any iterated roots.
The summary of the degree and the number of variables in polynomials $P$ that we found for all of the sectors is depicted in Figs~\ref{fig:geoms} and~\ref{fig:graphs}.
Overall, the maximal degree among these polynomials is 6, while the maximal number of the remaining unintegrated Baikov variables is 4.
These variables would have to be further integrated over two-dimensional surfaces or higher-dimensional hyper-surfaces for number of variables either 1 or more, respectively.
It is convenient to classify the special functions resulting from performing the integrals by the types of the underlying geometries defined by the algebraic varieties $w^2 = P(\vec{y})$ stemming from the roots in the integrand.
The special functions arising from the remaining univariate integrals are polylogarithms defined on a sphere~\cite{Chen:1977oja,Goncharov:1995ifj}, torus~\cite{Brown:2011wfj,Broedel:2017kkb,Bogner:2019lfa}, and genus two surface~\cite{Enriquez:2011np,Gorges:2023zgv,DHoker:2023vax,DHoker:2025szl}, for polynomial degree 2, 4, and 6, respectively.
The special functions arising from the remaining univariate integrals on hyper-surfaces are much less-undestood~\cite{Brown:2010bw,Bourjaily:2018ycu,Bourjaily:2018yfy,Bourjaily:2019hmc}.
It would be interesting to integrate them further to obtain a univariate integral form.
In comparison, all of the special functions appearing in one-loop Feynman integrals are polylogarithms defined on a sphere~\cite{Ellis:2007qk}.
In addition, we note three particular features of the summary in Figs~\ref{fig:geoms} and~\ref{fig:graphs}.
Firstly, the polynomials with degree 6 and either 1 or 2 variables stem purely from nonplanar sectors.
Secondly, for fully massless internal propagators, the only contribution to the polynomial with degree 6 and and 1 variable arises from sectors with more than 8 denominators, i.e. from the evanescent MIs which do not contribute to the finite part of any two-loop amplitude, see Sec.~\ref{sec:evan}.
Thirdly, for fully massless internal propagators, the polynomials with 4 variables and degree 4 or 6 do not appear when using the LBL Baikov representation instead of the standard Baikov one.

\begin{figure}[!h]
    \centering
    \includegraphics[width=0.8\textwidth]{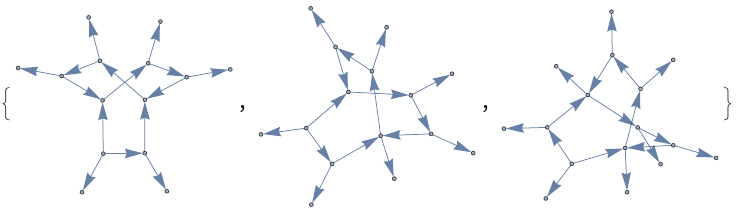}
    \caption{All topologies which always lead to iterated roots after one Baikov integration are evansescent.}
    \label{fig:iterRoot}
\end{figure}

All the sectors in Figs~\ref{fig:geoms} and~\ref{fig:graphs} have at least one variable in which the Baikov polynomial has degree 2, such that this variable can be chosen as the first integration variable.
The only sectors without this property are the 3 graphs in Fig.~\ref{fig:iterRoot}.
They all have Baikov polynomials with exactly 2 variables and degree 4 in each of these variables separately, as well as together, i.e. they have the form $\mathcal{B}(x_1,x_2) = c_1 x_1^4 + c_2 x_2^4 + c_3 x_1^3 x_2 + c_4 x_1 x_2^3 + c_5 x_1^2 x_2^2 + c_6 + \dots$ with kinematics-dependent coefficients  $c_i$.
Their residues in any of the two variables lead to iterated roots.
Importantly, all of these 3 sectors have 9 denominators, therefore, their MIs are evanescent, i.e. they will not contribute to the finite part of any two-loop amplitude, see Sec.~\ref{sec:evan}.
It is also worth pointing out that iterated roots can still appear for the graphs in Figs~\ref{fig:geoms} and~\ref{fig:graphs} after integrating out further Baikov variables, which is out of the scope of this work.

This concludes our analysis of the spectrum of special functions and the corresponding geometries appearing in any two-loop scattering amplitude in $d=4-2\ep$.
Note that putting an upper bound on this spectrum would require performing all the further possible residues. One would also have to account for the possible dualities between the higher genus curve and the Calabi-Yau geometries~\cite{Jockers:2024uan}.
Similarly to the finite basis discussion in Sec.~\ref{sec:fin}, the spectrum can simplify for specific physical scattering process with degenerate kinematics.
We note, that exploiting differential equations~\cite{Kotikov:1990kg,Gehrmann:1999as}, in addition to the Baikov polynomial method, for the analysis of MIs could illuminate further properties of the two-loop spectrum.
It would be especially important for the peculiar integrals that admit a canonical differential equation form~\cite{Henn:2013pwa} but do not integrate to Multiple Polylogarithms~\cite{Duhr:2020gdd}, e.g. due to the presence of not-simultaneously-rationalizable roots~\cite{Papathanasiou:2025stn}.
However, we leave this study for the future.

\section{Conclusions}
\label{sec:concl}

In this paper, we explored multiple implications of the existence of finite basis topologies for any two-loop amplitude in $d=4-2\ep$ found in Ref.~\cite{Bargiela:2024rul}.
Firstly, we found a functionally distinct finite basis of all Feynman integrals at two loops.
For all off-shell legs, it consists of 347 MIs for fully massive internal lines, and 274 MIs for massless.
We present this finite basis in Tab.~\ref{tab:MIs} and in the ancillary files.
Secondly, we have shown on a numerical probe that all the two-loop MIs with more than 8 denominators do not contribute to the finite part of any two-loop amplitude.
To this end, we constructed an example basis of IR and UV finite MIs defined in $d+2=6-2\ep$ dimensions with linear numerators.
Then, we explicitly found an overall $\ep$ factor in front of this basis after the IBP reduction of the MIs in Tab.~\ref{tab:MIs} onto this basis.
Thirdly, we elaborated on the applications of the tHV decomposition to numerically evaluate CDR integrals.
On a two-loop six-point example, we removed redundant CDR numerators to improve the \texttt{AMFlow} performance.
On a one-loop high-multiplicity example, we removed redundant CDR denominators to improve the \texttt{DCT} performance.
Finally, we analyze the spectrum of special functions and the corresponding geometries appearing in any two-loop scattering amplitude in $d=4-2\ep$.
We computed the Baikov polynomial in each subsector of the finite basis topologies and integrated out one of the Baikov variables.
As a result, we observed that remaining integrands involve square roots of polynomials with at most degree 6 and 4 variables.
This corresponds to sphere, torus, and genus two geometries, as well as more involved hyper-surfaces.

Our analysis induces natural steps to pursue in the future.
Firstly, a lot of the finite basis MIs are still awaiting to be calculated explicitly.
For the phenomenologically motivated kinematics, there has been a major effort in this direction, for two-loop four-point~\cite{Chaubey:2024ckh}, five-point~\cite{Chicherin:2020oor,Chicherin:2021dyp,Badger:2024fgb,FebresCordero:2023pww}, and six-point~\cite{Abreu:2024fei,Henn:2025xrc} scattering.
Secondly, it would be important to verify analytically that all of the two-loop MIs with more than 8 denominators can be chosen to be evanescent.
For the massless on-shell double-pentagon and hexa-box, it has been already shown in Ref.~\cite{Abreu:2024fei}.
Thirdly, the tHV decomposition will improve the performance of the IBP reduction for both integrals and amplitudes, and the associated numerical evaluation, in new high-multiplicity processes.
Indeed, it has been already used for reducing the complexity of the numerical evaluation of the massless on-shell double-pentagon MIs in Ref.~\cite{Henn:2025xrc}.
Finally, the classification of the special functions and the corresponding geometries would help in constructing a more efficient functional ansatzes for Feynman integrals and scattering amplitudes in the bootstrap approach~\cite{Dixon:2011pw,Chicherin:2017dob,Hannesdottir:2024hke}.
In addition, it would be beneficial to further constrain the spectrum, e.g. by simplifying some of the hyper-surfaces stemming from the 30 corresponding    graphs in Figs~\ref{fig:geoms} and~\ref{fig:graphs}.

\section*{Acknowledgements}

We thank R. Marzucca for valuable comments on this paper draft.
We are grateful to K. Sch{\"o}nwald, V. Sotnikov, and S. Zoia for important comments on the evanescent integrals.
We also thank G. Fontana, H. Frellesvig, F. Lange, R. Morales, S. P{\"o}gel, and M. Wilhelm for interesting discussions.
This research was supported by the Swiss National Science Foundation (SNF) under contract 200020-204200 and by the European Research Council (ERC) under the European Union's Horizon 2020 research and innovation programme grant agreement 101019620 (ERC Advanced Grant TOPUP).

\newpage

\appendix
\section{Summary of the properties of the two-loop finite basis}
\label{app:1}

In this appendix, we provide graphical summaries referred to in the previous sections of this paper.

\begin{figure}[!h]
    \centering
    \begin{subfigure}[b]{0.85\textwidth}
         \centering
         \includegraphics[width=0.99\textwidth]{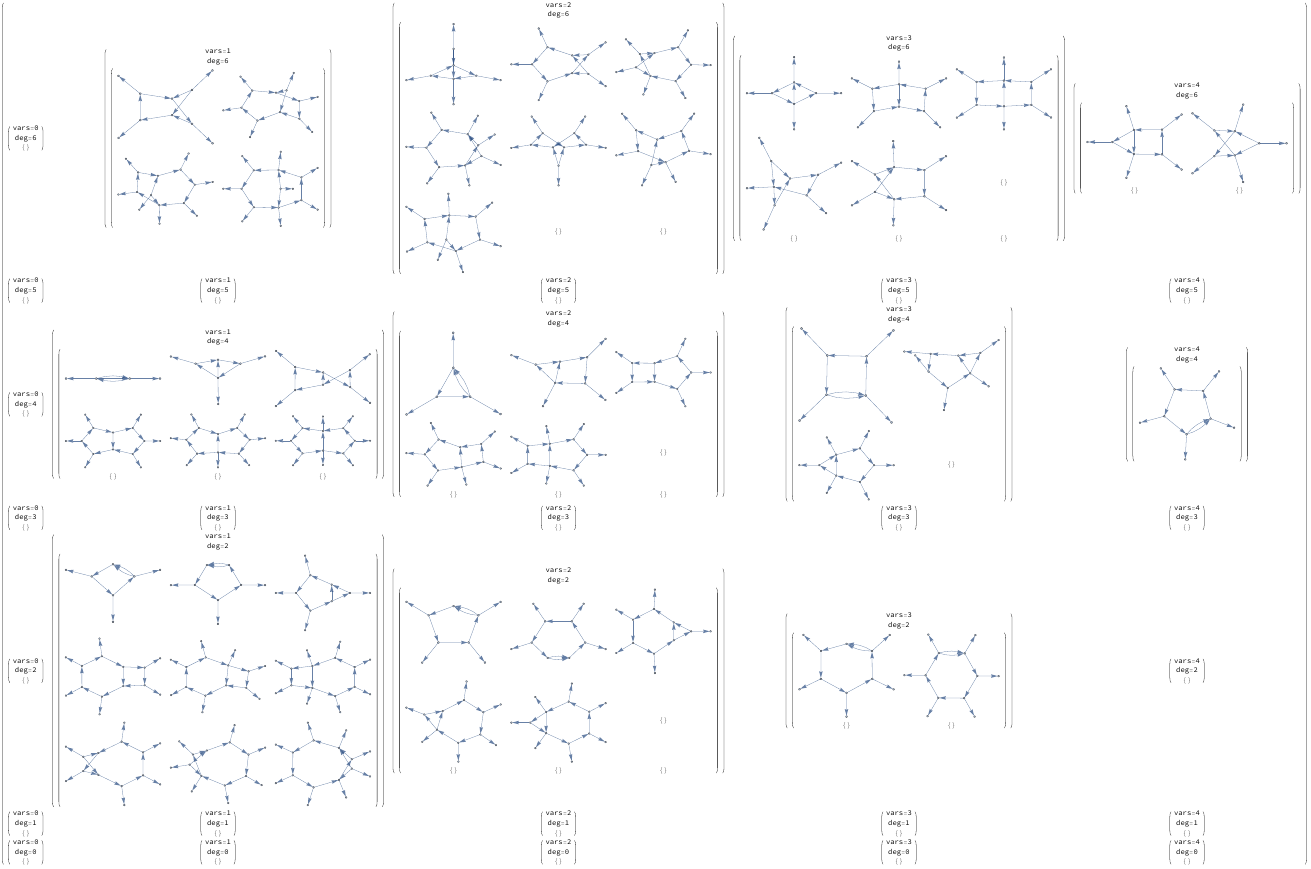}
         \caption{massive}
     \end{subfigure}
     \begin{subfigure}[b]{0.85\textwidth}
         \centering
         \includegraphics[width=0.99\textwidth]{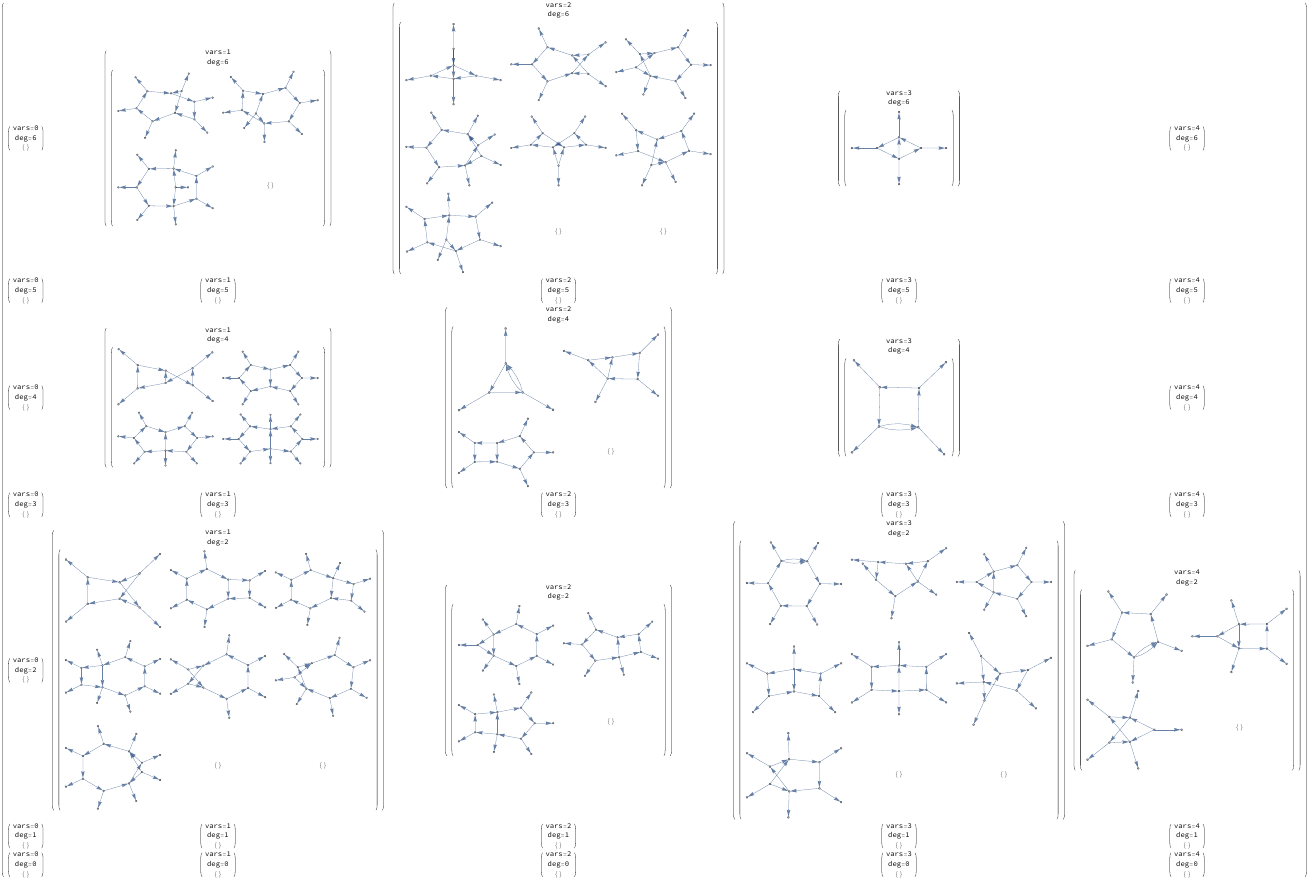}
         \caption{massless}
    \end{subfigure}
     \caption{Same plot as in Fig.~\ref{fig:geoms} but with associated Feynman graphs for all massive and massless propagators, respectively.}
     \label{fig:graphs}
\end{figure}

\newpage

\begin{figure}[!h]
    \centering
    \includegraphics[width=0.574\textwidth]{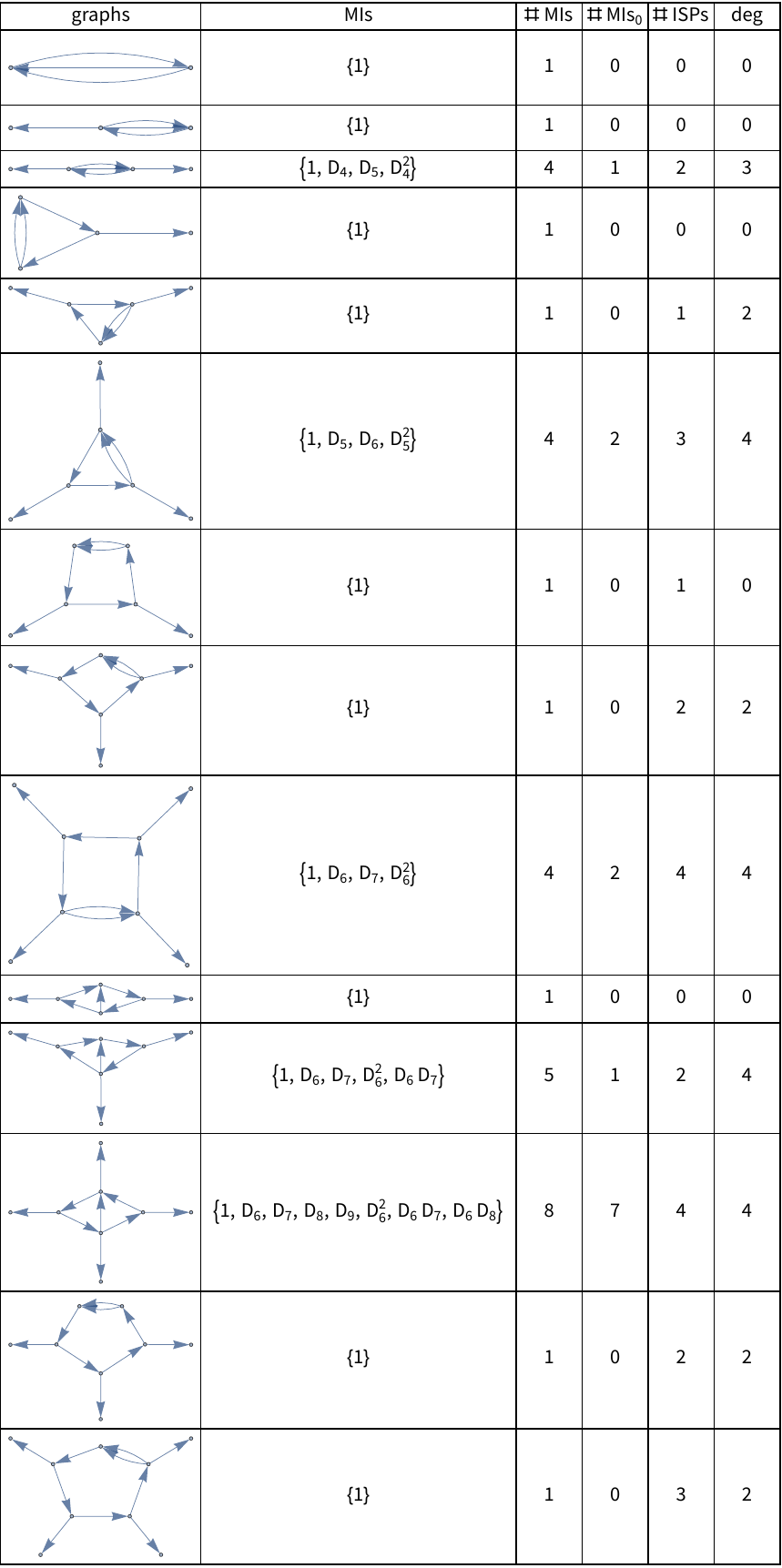}
    \includegraphics[width=0.39\textwidth]{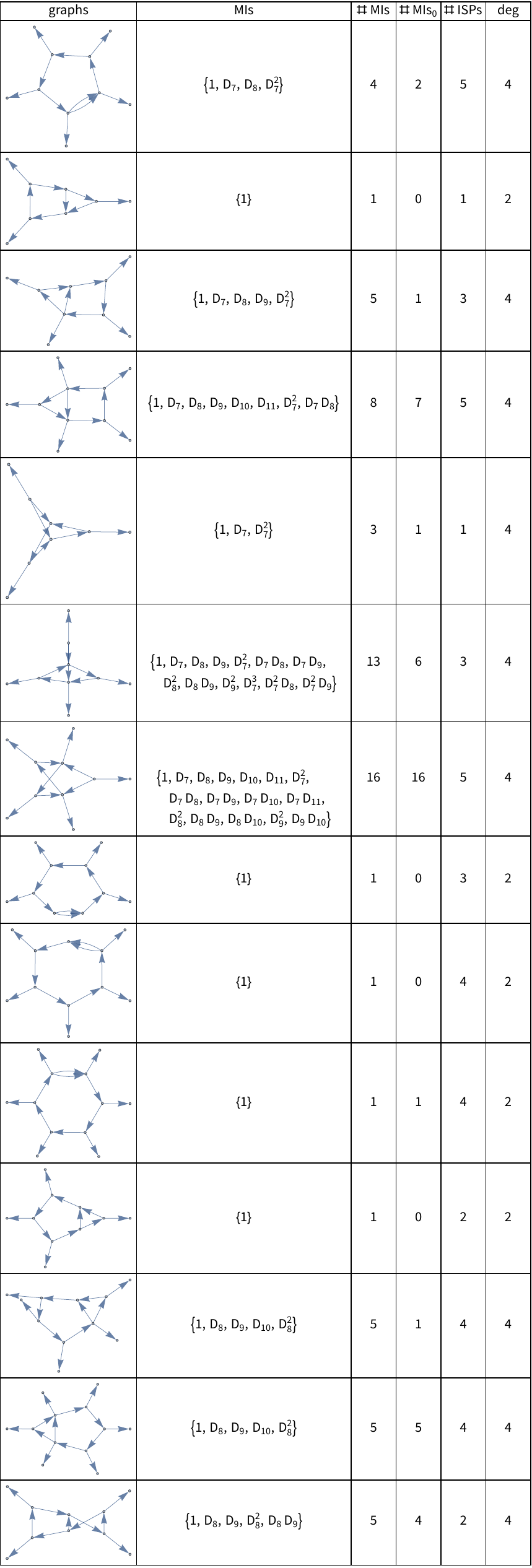}
\end{figure}

\newpage

\begin{figure}[!h]
    \centering
    \includegraphics[width=0.425\textwidth]{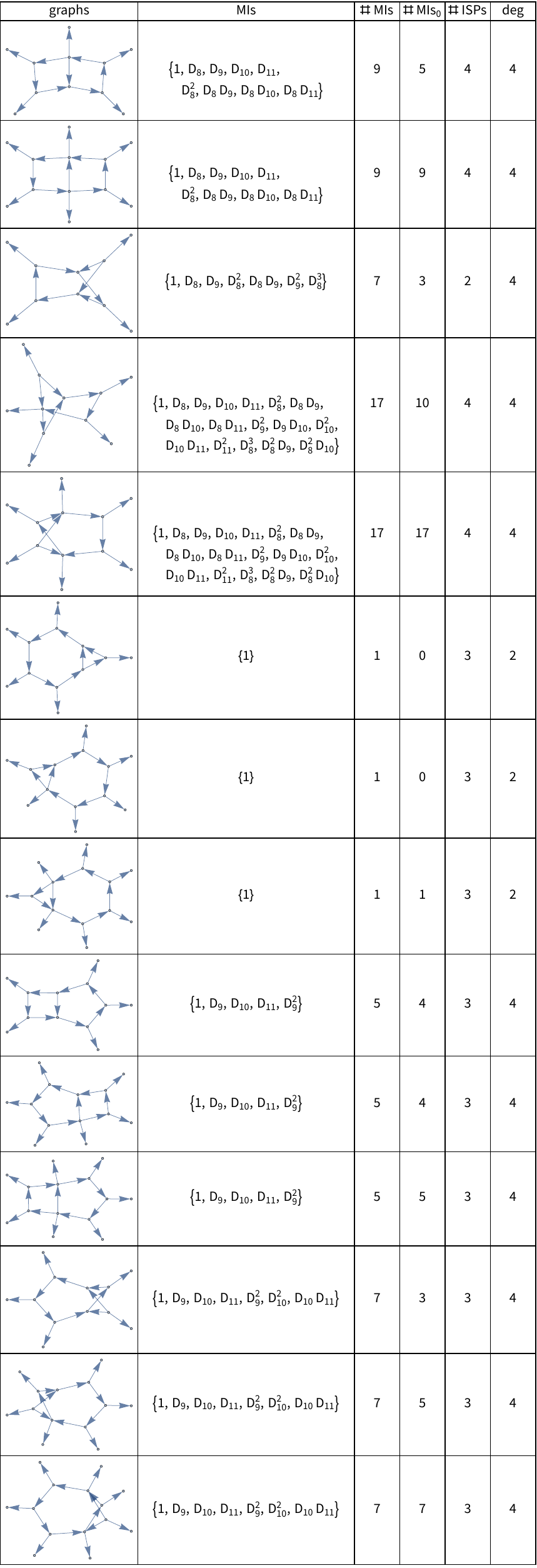}
    \includegraphics[width=0.45\textwidth]{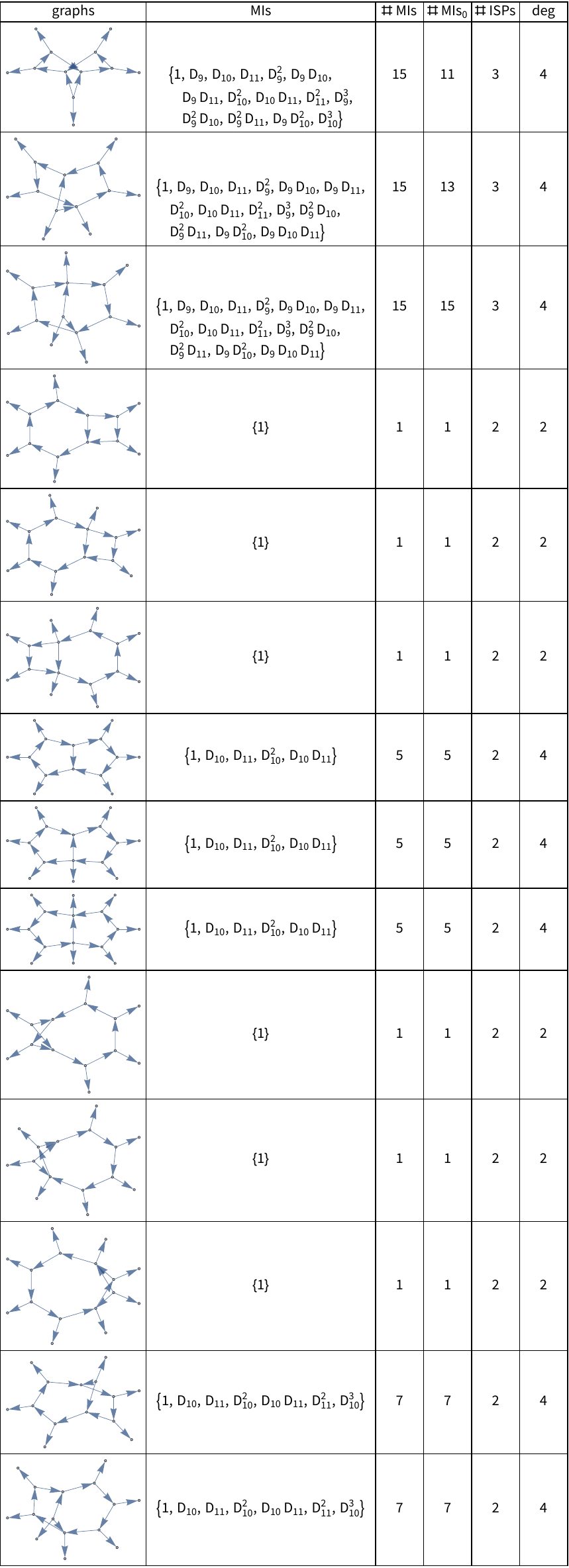}
\end{figure}

\newpage

\begin{table}[!h]
    \centering
    \includegraphics[width=0.437\textwidth]{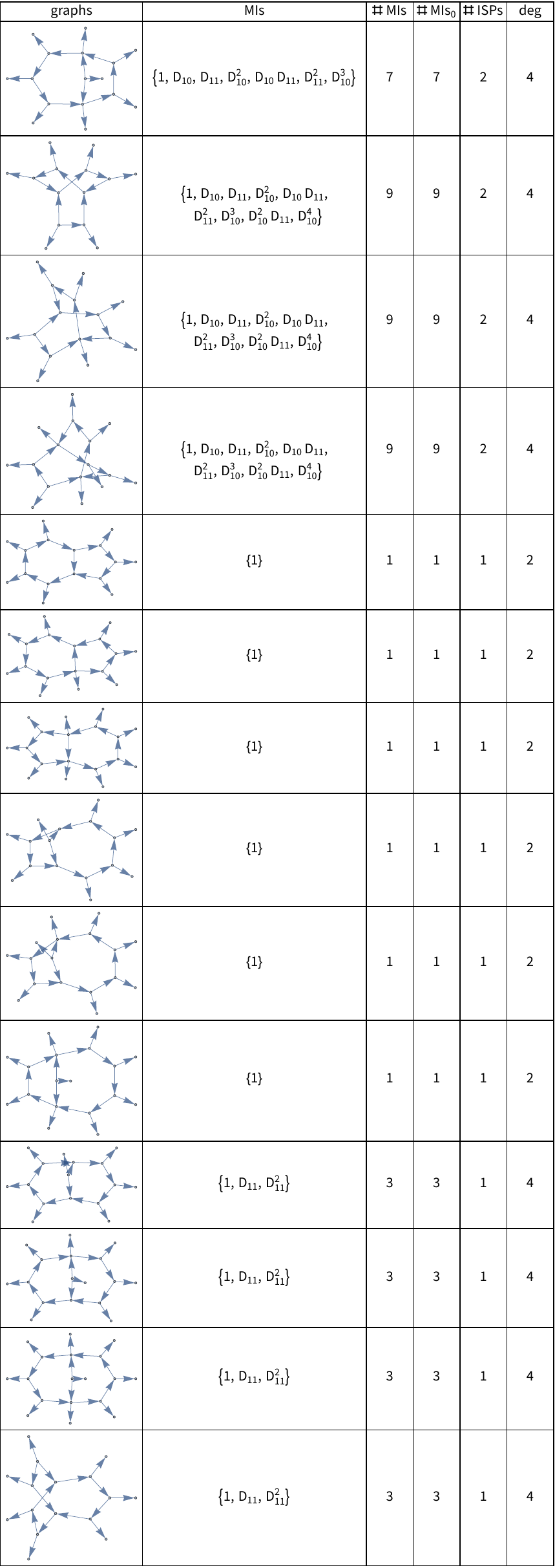}
    \includegraphics[width=0.45\textwidth]{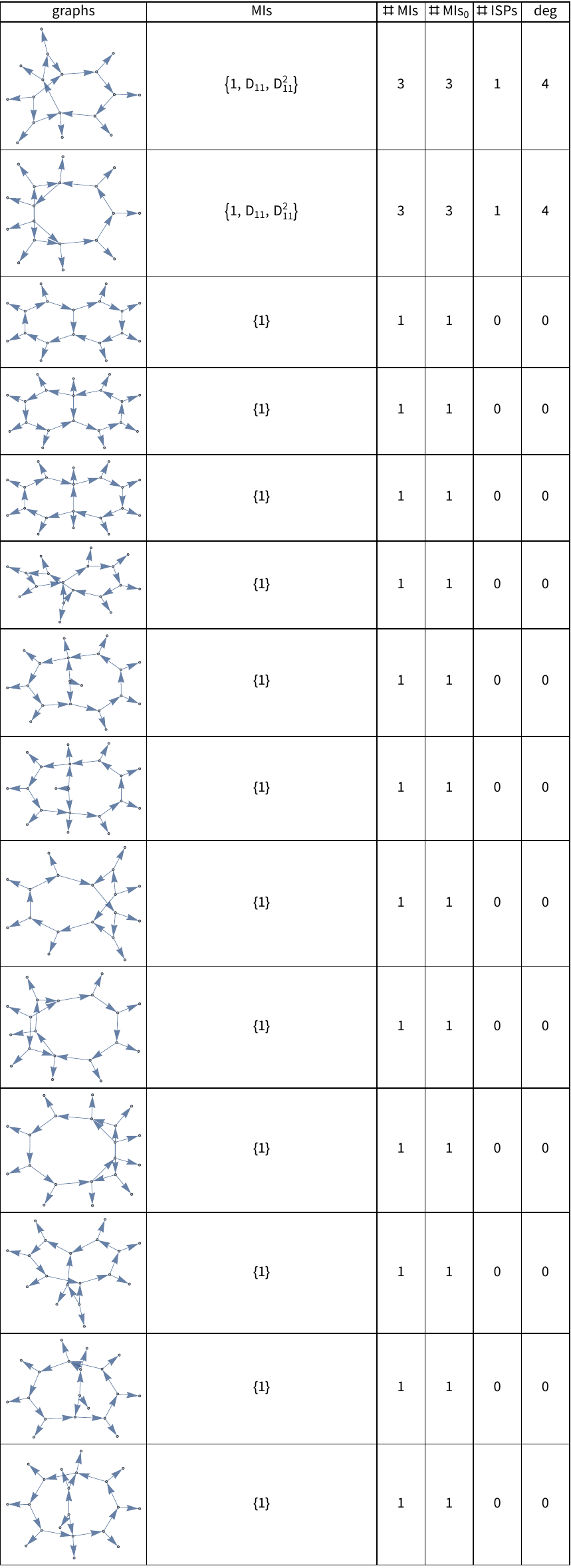}
    \caption{The graph, numerators of MIs, number of MIs in both all internally massive and massless cases, number of ISPs, and the degree of the Baikov polynomial for all the two-loop finite basis Master Integrals.}
\label{tab:MIs}
\end{table}

\newpage

\bibliographystyle{JHEP}
\bibliography{references}

\end{document}